\documentclass[a4paper, amsfonts, amssymb, amsmath, reprint, showkeys, nofootinbib, twoside]{revtex4-2}
\usepackage[english]{babel}
\usepackage[utf8]{inputenc}
\usepackage{appendix}

\usepackage[plain]{algorithm}
\usepackage{amsmath}
\usepackage{amssymb}
\usepackage{bm}
\usepackage[nolabel]{showlabels}
\usepackage[ISO=false]{diffcoeff}
\usepackage{listings}
\usepackage{graphicx}
\usepackage[utf8]{inputenc}
\usepackage{tablefootnote}
\usepackage{threeparttable}
\usepackage{makecell}
\usepackage{xcolor}
\definecolor{mydarkgray}{gray}{0.2}

\usepackage{color}

\usepackage{graphics}

\newcommand{\avg}[1]{\left\langle{#1}\right\rangle}
\newcommand{\qt}{{q_{\theta}}}
\newcommand{\qtt}{{\tilde{q}_{\theta}}}

\renewcommand{\v}[1]{\mathbf{#1}}

\newcommand{\wh}{\hat{w}}

\usepackage{graphicx}
\usepackage{dcolumn}
\usepackage{bm}
\usepackage{textcomp}
\usepackage{color}
\usepackage{amsmath}
\usepackage[ISO=false]{diffcoeff}

\begin{document}

\title{Mutual information of spin systems from autoregressive neural networks}

\author{Piotr Białas}
\email{piotr.bialas@uj.edu.pl}
\affiliation{Institute of Applied Computer Science, Jagiellonian University, ul. \L ojasiewicza 11, 30-348 Krak\'ow, Poland}
\author{Piotr Korcyl}
\email{piotr.korcyl@uj.edu.pl}
\author{Tomasz Stebel}
\email{tomasz.stebel@uj.edu.pl}
\affiliation{Institute of Theoretical Physics, Jagiellonian University, ul. \L ojasiewicza 11, 30-348 Krak\'ow, Poland}

\date{\today}

\begin{abstract}
We describe a new direct method to estimate bipartite mutual information of a classical spin system based on Monte Carlo sampling enhanced by autoregressive neural networks. It allows studying arbitrary geometries of subsystems and can be generalized to classical field theories. We demonstrate it on the Ising model for four partitionings, including a multiply-connected even-odd division. We show that the area law is satisfied for temperatures away from the critical temperature: the constant term is universal, whereas the proportionality coefficient is different for the even-odd partitioning.

\end{abstract}

\keywords{Mutual Information, Hierarchical Autoregressive Neural Networks, Monte Carlo simulations, Ising model}

\maketitle

\section{Introduction}

The discovery of topological order in quantum many-body systems \cite{Zeng_2019} initiated a very fruitful exchange of ideas between solid-state physics and information theory. Many new theoretical tools developed to quantitatively describe the flow of information or the amount of information shared by different parts of the total system have been employed in the studies of physical systems \cite{Okunishi_2022}. Among these tools, quantum entanglement entropy or mutual information and their various alternatives were found to be particularly useful \cite{2004JSMTE..06..002C,2008PhRvA..78c2319A,RevModPhys.82.277,2011PhRvL.106t1601A,2013PhRvL.110i1602B,Eisert_2013,2018PhRvL.120t0602G,2020JSMTE2020g3101C}. With their help, it was understood that some new phases of matter respect the same set of symmetries but differ in long-range correlations quantified by bipartite, tripartite, or higher information-theoretic measures such as mutual information \cite{Kitaev:2005dm,Levin_2006}. Turning this fact around, it is expected that calculating mutual information can provide hints about the topological phase of the system, playing a role similar to order parameters in the usual Landau picture of phase transitions (see for example \cite{Iaconis_2013,PhysRevB.101.085136}). Indeed, such quantities are not only useful for theoretical understanding but are also measurable observables in experiments. For example, in Ref.~\cite{Tajik:2022ycs} the mutual information in a quantum spin chain was measured demonstrating the area law \cite{Wolf:2007tdq} governing the scaling of mutual information with the volume of the bipartite partition. The law \cite{PhysRevLett.71.666} claims that the mutual information or entanglement entropy in the thermodynamic limit scale with the boundary area separating the two parts of the system, instead of the volume, as for the other extensive properties. The increased interest comes from the fact that the black hole entropy was shown a long time ago \cite{PhysRevD.7.2333,Hawking:1975vcx} to follow a similar law: the black hole entropy depends only on its surface and not on the interior. To a large surprise, this connection was recently made explicit with the realization that a certain quantum spin model in 0 spatial dimensions called the SYK model \cite{Sachdev_1993} is dual to a black hole via the gravity/gauge duality \cite{Maldacena:1997re}. From this perspective, quantum entanglement entropy or quantum mutual information which can be defined and calculated on both sides of this correspondence became even more attractive from the theoretical point of view. 
Therefore there is a strong pressure to develop computational tools able to estimate such quantities, which in itself is, however, very difficult in the general case. 

Although an important part of Condense Matter physics \cite{Wolf:2007tdq,2008RvMP...80..517A,Stephan_2009,Alcaraz_2013,Alcaraz_2014} aims at studying entanglement entropy and quantum mutual information in quantum spin systems, already in classical spin systems the Shannon and Renyi entropies were found to follow the area law \cite{St_phan_2010}. With the interface boundary sizes up to $64$ spins the coefficients of the area law were precisely determined and their universality was verified using different lattice shapes. Later, the mutual information of two halves of the classical Ising model on an infinitely long cylinder was calculated using the transfer matrix approach \cite{Wilms_2011}. Both quantities were discussed with more precision in \cite{Lau_2013} using a different method, called the bond propagation algorithm. Similarities between quantum entanglement entropy and classical mutual information were studied in Ref.~\cite{PhysRevB.106.214303}. 

Recent developments in machine learning algorithms open new possibilities to study information theory observables. Quantum entanglement entropy can be calculated using an approximation of the ground state provided by neural networks. For example, in Refs.~\cite{2020PhRvR...2b3358H,2020PhRvA.102f2413W} autoregressive architectures were used to calculate variational Renyi entropies for 1D and 2D Ising and Heisenberg models. In Ref.~\cite{2020PNAS..11730234N}  mutual information of classical spin systems was calculated using the method called machine-learning
iterative calculation of entropy (MICE). It is based on the idea of Ref.~\cite{pmlr-v80-belghazi18a} of exploiting Donsker-Varadhan representation of Kullback-Leibler (KL) divergence. We discuss further the MICE method in Appendix \ref{app_MICE_comp}.

In this work, we propose a new method to directly estimate classical bipartite mutual information. It is based on the incorporation of machine learning techniques into Monte Carlo simulation algorithms. From the theoretical point of view, the algorithm we use belongs to the class of Metropolized Independent Sampling \cite{Liu} algorithms. Therefore, our calculations are stochastic and provably exact within their statistical uncertainties, assuming that all the modes of the target distribution are probed (no mode collapse). The main advantage of the method is its flexibility, allowing its application: to any geometry of the partitioning, to any statistical system with a finite number of degrees of freedom, in an arbitrary number of space dimensions, provided that the target probability can be effectively trained ({\it e.g.} with no mode collapse).

\section{Method}

\subsection{Mutual information (MI) for spin systems}

We consider a classical 
system of spins which we divide into two arbitrary parts $A, B$. In this case, the Shannon mutual information is defined as 
\begin{equation}
   I = \sum_{\v a, \v b}p(\v a, \v b)\log \frac{p(\v a, \v b)}{p(\v a)p(\v b)},
    %
    \label{eq. MI}
\end{equation}
where a particular configuration $\v s$ of the full model has parts $\v a$ and $\v b$, and where the Boltzmann probability distribution of states, depending on inverse temperature $\beta$, is given by (we omit the explicit dependence of $Z$ on $\beta$),
\begin{equation}
p(\v a, \v b) = \frac{1}{Z} e^{-\beta E(\v a, \v b)}, \quad Z = \sum_{\v a,\v b} e^{-\beta E(\v a, \v b)}
\label{eq. p}
\end{equation}
and
\begin{equation}\label{eq:pa-pb}
p(\v a) =  \sum_{\v b } p(\v a, \v b), \ \ p(\v b) = \sum_{\v a} p(\v a, \v b)
\end{equation}
are probability distributions of subsystems. We shall use the same symbol $p$ for all probability distributions defined on different state spaces distinguishing them by the arguments. In the above expressions, the summation was performed over all configurations of subsystems $A$ or $B$. Inserting Eq.~\eqref{eq. p} and Eq.~\eqref{eq:pa-pb} into Eq.~\eqref{eq. MI} we obtain
\begin{multline}
I = \log Z - \sum_{\v a ,\v b }p(\v a, \v b)\Big[ \beta E(\v a, \v b) 
+\\+ 
\log Z(\v a) + \log Z(\v b) \Big],
\label{eq. MI2}
\end{multline}
where
$Z(\v a ) = \sum_{\v b } e^{-\beta E(\v a, \v b)}$ and $Z(\v b ) = \sum_{\v a } e^{-\beta E(\v a, \v b)}$.
Please note that in typical Monte Carlo approaches the partition functions $Z$, $Z(\v a)$, and $Z(\v b)$ are not available.

\subsection{Neural Importance Sampling for MI}

 Below we argue that $I$ can be obtained from a Monte Carlo simulation enhanced with autoregressive neural networks (ANNs). It was recently shown that ANNs can be used to approximate the Boltzmann probability distribution $p(\v a, \v b)$ for spin systems and provide a mean to sample from this approximate distribution \cite{2019PhRvL.122h0602W,Bialas:2021bei,Bialas:2022qbs,Bialas:2022bdl}.  Let us call this approximate distribution $\qt(\v a, \v b)$, $\theta$ stands here for the parameters of the neural network that are to be tuned so that $\qt$ is as close to $p$ as possible under an appropriate measure, typically the backward Kullback-Leibler divergence
\begin{align}
    D_\textrm{KL} (q_\theta | p) &= \sum_{\v a, \v b} q_\theta(\v a, \v b) \, \log \left(\frac{q_\theta(\v a,\v b)}{p(\v a,\v b)}\right).
    \label{eq:KL_loss} 
\end{align}

The formula Eq.~\eqref{eq. MI2} can be rewritten in terms of  averages with respect to the distribution $\qt$
\begin{multline}
I = \log Z -\frac{1}{Z}\beta\avg{\wh(\v a, \v b)E(\v a, \v b)}_{\qt(\v a, \v b)}+\\
+\frac{1}{Z}\avg{\wh(\v a, \v b)\log Z(\v a)}_{\qt(\v a, \v b)} +\\+ \frac{1}{Z}\avg{\wh(\v a, \v b)\log Z(\v b)}_{\qt(\v a, \v b)},
\label{eq. MI4}
\end{multline}
where the importance ratios are defined as
\begin{equation}
    \hat w(\v a, \v b) = \frac{e^{-\beta E(\v a, \v b)}}{q_{\theta}(\v a, \v b)}.
    \label{importance_ratio_def}
\end{equation}

The crucial feature of ANN-enhanced Monte Carlo is that contrary to standard Monte Carlo, we can estimate directly the partition functions $Z$, $Z(\v a)$ and $Z(\v b)$ (see 
Appendix \ref{ap. systematics}
for details). 
In this way, we have expressed the mutual information $I$ only through averages with respect to the distribution $\qt$. It can be now estimated by sampling from this distribution. The procedure of sampling configurations from approximate probability distribution provided by neural networks together with reweighting observables with importance ratios was proposed in Ref.~\cite{2020PhRvE.101b3304N} and named Neural Importance Sampling (NIS). This paper reports on the first application of this technique to information theory observables. 

Autoregressive neural networks rely on the product rule {\em i.e.} factorization of $\qt$ into the product of conditional probabilities
\begin{equation}
\label{eq:factorisation}
    q_{\theta}(\v a, \v b)  =  \prod_{i=1}^{L^2} q_{\theta}(s^i|s^1,s^2,\dots,s^{i-1}).
\end{equation}
Due to the fact that the labeling of spins in Eq.~\eqref{eq:factorisation} is arbitrary, we can choose it in such a way that we first enumerate all spins from part $A$, $\v a=(s^1, s^2,\ldots,s^{n_A})$ and only afterward all spins from part $B$, $\v b = (s^{n_A+1},\ s^{n_A+2},\ldots,s ^{n_A+n_B})$. We then obtain 
\begin{equation}
    \qt(\v s) \equiv \qt(\v a, \v b)  = \qt(\v a) \qt(\v b|\v a),
\end{equation}
with 
\begin{equation}
\qt(\v a) =  \prod_{i=1}^{n_A}\qt(s^i|s^1,s^2,\dots,s^{i-1}),
\end{equation}
and
\begin{multline}
\qt(\v b | \v a) = 
\prod_{i=1}^{n_B}
\qt(s^{n_A+i}|s^{n_A+1},s^{n_A+2},\dots,s^{n_A+i-1}, \v a).
\label{eq. conditional probability}
\end{multline}
These features of ANN stand behind the fact that we can readily estimate $\log Z(\v a)$ and $\log Z(\v b)$ required by Eq.~\eqref{eq. MI4} directly. In this respect, the ANN approach differs from {\em normalizing flows} used to approximate continuous probability distributions and employed recently in the context of lattice field theory \cite{PhysRevD.100.034515,Nicoli:2020njz,2021arXiv210108176A,DelDebbio:2021qwf, Albandea:2021kwe, Bialas:2022iro, Abbott:2022hkm,Bialas:2023fyj}. There, the probability is calculated for the whole field configuration at once and the conditional probability $\qt(\v b|\v a)$ as well the marginal distribution $\qt(\v a)$ are not so easily available.

\section{Results}

\begin{figure}
\begin{center}
\includegraphics[width=0.5\textwidth]{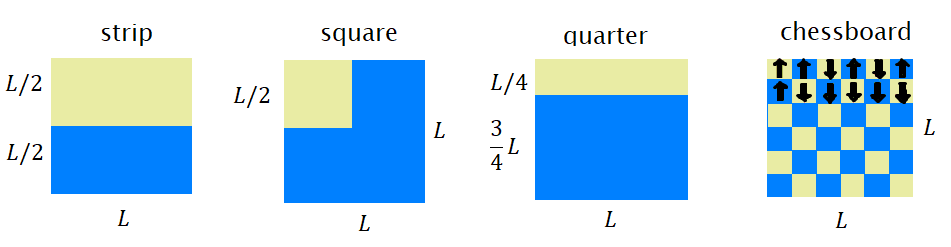}
\vspace{-0.5cm}
\caption{Four considered partitioning geometries. Periodic boundary conditions are applied. The small blocks in the chessboard partitioning represent single spins.\label{partition_geom}}
\end{center}
\vspace{-0.5cm}
\end{figure}

 We leave the technical details of the evaluation of the terms in Eq.~\eqref{eq. MI4} to the 
Appendix \ref{ap. systematics}
and now provide an example of its application. We demonstrate it on the Ising model on a periodic $L \times L$ lattice, with ferromagnetic, nearest-neighbor interactions, defined by the Hamiltonian
\begin{equation}
  E(\v a, \v b)= -\sum_{\langle i,j \rangle} s^i s^j, 
\end{equation}
where $s^i \in \{1, -1\}$. We consider the following divisions and respective mutual information observables:
\begin{itemize}
    \item "strip" geometry -- the system is divided into equal rectangular subsystems;
    \item "square" geometry -- subsystem $A$ is the square of size $\frac{1}{2}L \times \frac{1}{2}L$;
    \item  "quarter" geometry -- $A$ is rectangular of size $\frac{1}{4}L \times L$.
\end{itemize}
 We show them schematically in the three sketches on the left in Fig.~\ref{partition_geom}. We note that all of them have the same length of the border between the subsystems, i.e. $2L$ (as we used periodic boundary conditions) although may they differ in their volumes. 
In the discussion below, we refer to those three partitionings as "block" partitionings. In addition, we also consider a division, called
\begin{itemize}
    \item "chessboard" partitioning, where the system is divided using even-odd labeling of spins. 
\end{itemize}
In this case, the boundary between spins is $L^2$, i.e. every spin of the system is at the boundary between parts $A$ and $B$. Note that in the chessboard partitioning the subsystems are not simply connected, as is usually considered in the Literature.

We investigate a wide range of system sizes, reaching $L=66$ for $\beta > 0.3$ and  $L=130$ for $\beta \le 0.3$\footnote{The reason behind this choice is that for larger $\beta$ training of the neural networks is less efficient and for these temperatures we did not reach sufficient quality of training. This could be in principle done using additional tricks, like {\it e.g.} pretraining which we used for simulations of the Potts model~\cite{Bialas:2022bdl}.}. We combine results obtained  by the Variational Autoregressive Network (VAN) approach of Ref.~\cite{2019PhRvL.122h0602W} and our recently proposed modification called Hierarchical Autoregressive Network (HAN) algorithm \cite{Bialas:2022qbs}. Both approaches are applied to several divisions, as summarized in Table~\ref{tab: L values}.
We describe the details of the VAN and HAN architectures and the quality of their training in 
Appendix \ref{ap. neural nets}. In Appendix \ref{ap. chessboard} we discuss the details of mutual information calculation for the chessboard partitioning. Simulations are performed at 13 values of the inverse temperature: from $0.1$ to $0.4$ with a step $0.05$, $0.44$, and from $0.5$ to $0.9$ with a step $0.1$.
For each $\beta$ we collect the statistics of at least $10^6$ configurations and estimated the statistical uncertainties using the jackknife resampling method. In all cases, the relative total uncertainty of $I$ combining the systematic and statistical uncertainties is of the order of $1$ \textperthousand, or smaller 
-- see appendix \ref{ap. systematics} (in particular the discussion around Eq.~(\ref{bias_INM})).

\begin{table}
    \centering
    \begin{tabular}{|c|c|c|}
    \hline
    Partitioning & VAN & HAN \\
    \hline 
    strip & 8,12,16, & 10,18,34, \\
     & 20,24,28 & 66,130(for $\beta\le 0.3$) \\
    \hline 
    square &  & 10,18,34, \\
     &  & 66,130(for $\beta\le 0.3$) \\
    \hline 
    quarter & 8,12,16, &\\
     & 20,24,28 & \\
    \hline
    chessboard & 8,12,16,20, & \\
     & 24,28,32& \\
    \hline
    \end{tabular}
\caption{The system sizes $L$ which were considered for given partitioning type. Note that HAN requires total system size $L=2^n+2$, where $n$ would be the number of levels in the hierarchy.}
\label{tab: L values}    
\end{table}

\begin{figure}
\begin{center}
\vspace{-0.75cm}
\includegraphics[width=0.495\textwidth]{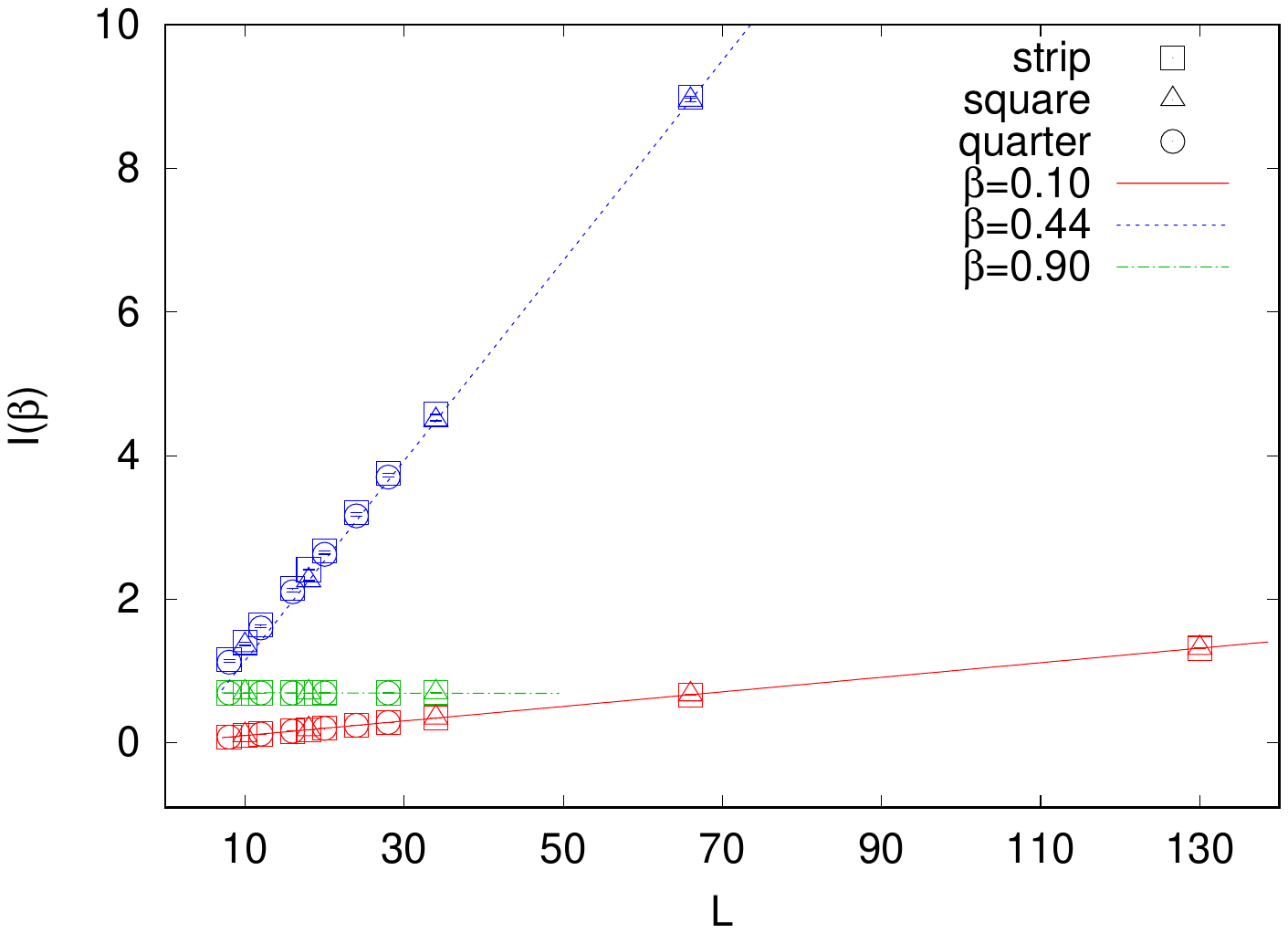}
\vspace{-1cm}
\caption{General dependence of $I$ on $L$ for various geometries and three representative inverse temperatures: $\beta=0.10$ in the disordered phase, $\beta=0.44$ close to the phase transition and $\beta=0.90$ in the ordered phase. Lines correspond to attempted fits using the area law Ansatz.\label{fig. scaling}}
\end{center}
\vspace{-0.75cm}
\end{figure}
\begin{figure}
\begin{center}
\includegraphics[width=0.495\textwidth]{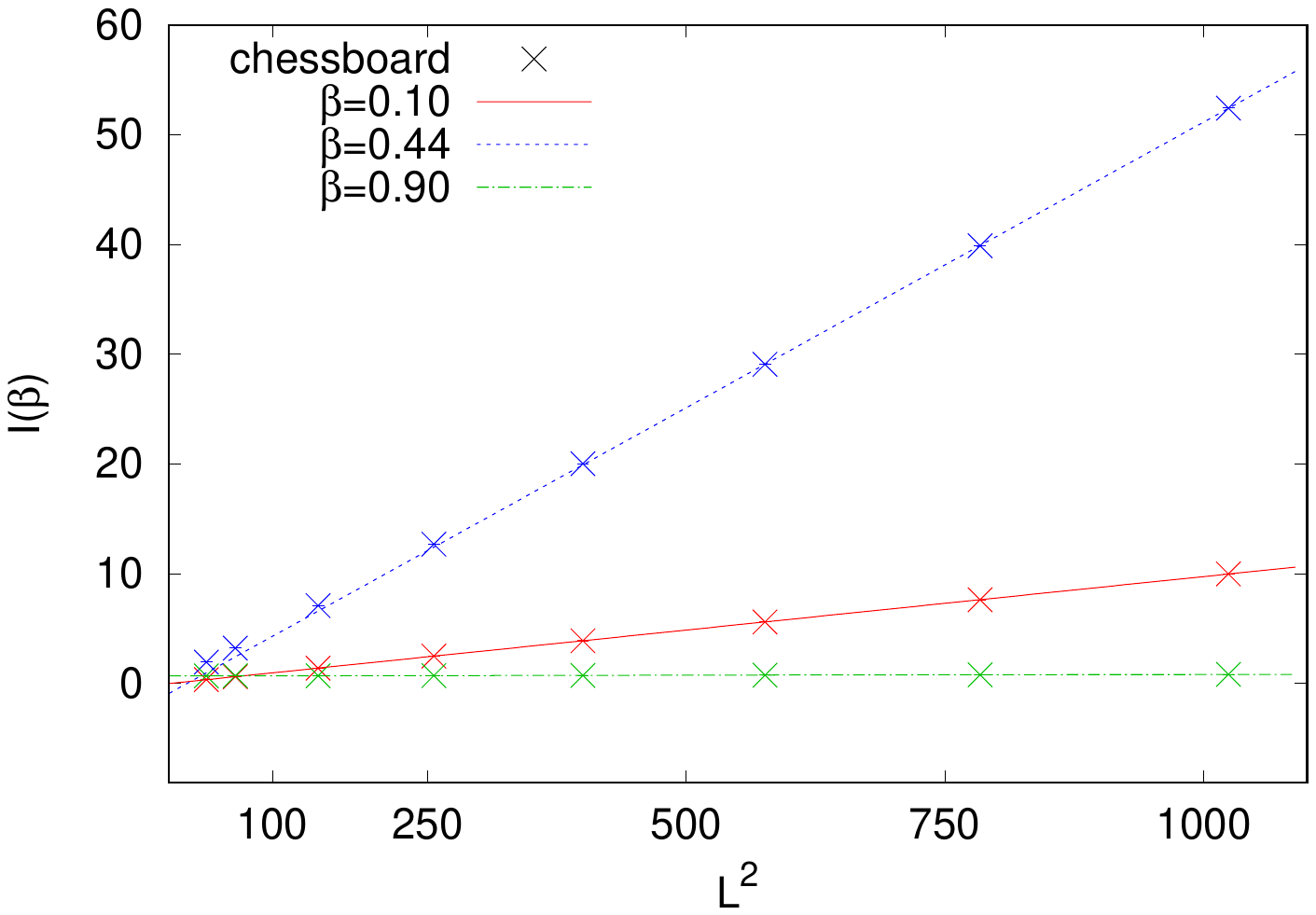}
\vspace{-1.25cm}
\caption{General dependence of $I$ on $L^2$ for the chessboard partitioning at three inverse temperatures: $\beta=0.10$, $\beta=0.44$ and $\beta=0.90$. Lines correspond to attempted fits using the area law Ansatz. Statistical uncertainties are shown but are much smaller than the symbol size.\label{fig. scaling chessboard}}
\end{center}
\vspace{-0.75cm}
\end{figure}

\subsection{Area law}

In Fig.~\ref{fig. scaling} we show $I$ as a function of $L$ for the three block partitionings and for three representative inverse temperatures $\beta$: 0.1 (high-temperature regime), 0.44 (very close to critical temperature $\beta_c=1/2 \ln(1+\sqrt 2)\approx 0.44069$), and 0.9 (low-temperature regime). We clearly observe a linear dependence on the system size with a strongly $\beta$-dependent slope and intercept. Similar behavior can be observed for the chessboard partitioning when one plots $I$ as the function of $L^2$ - see Fig.~\ref{fig. scaling chessboard}. The emerging picture, supporting the conjectured area law, is that $I$ can be written in a compact form as
\begin{equation}
I_{geom}(\beta,L) = \alpha_{geom}(\beta) B(L) + r_{geom}(\beta),
\label{eq. MI scaling}
\end{equation}
where the variable $B(L)$ corresponds to the length of the boundary between the two parts $A$ and $B$: $B(L)=L^2$ for the chessboard partitioning and $B(L)=2L$ for other partitionings considered in this work.
Eq.~\eqref{eq. MI scaling} has two parameters, $\alpha$ and $r$, which, as we denoted, depend on $\beta$ and can differ for different partitioning geometries.

\begin{figure}
\begin{center}
\includegraphics[width=0.495\textwidth]{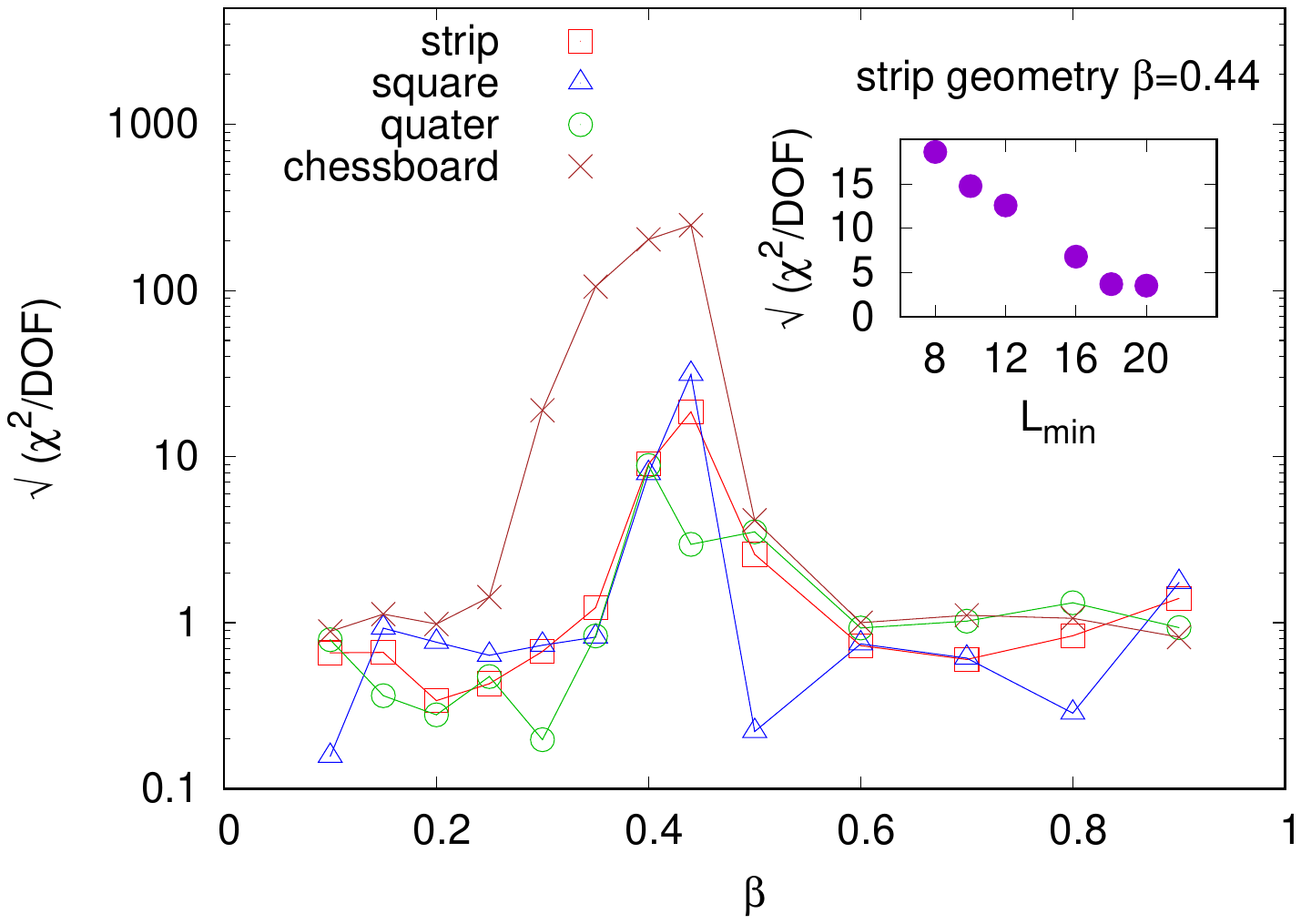}
\vspace{-1.0cm}
\caption{Quality of fits involving all available data points and assuming the area law functional form of the mutual information for the four geometries shown in Figure \ref{partition_geom}: $\sqrt{\chi^2/\textrm{DOF}}$ plotted as a function of the inverse temperature. In the inset, we show how the fit improves as we restrict the system sizes included in the fit to be bigger or equal to $L_{\textrm{min}}$. \label{fig. chi2}}
\end{center}
\vspace{-0.75cm}
\end{figure}

We expect that Eq.~\eqref{eq. MI scaling} is valid as long as finite volume effects in $I$ are small. 
In general, they may depend on three length scales present in the system: the size of the whole system $L$, the size of the smallest subsystem, and correlation length $\xi$ determined by the temperature of the system. Finite volume effects should vanish when $\xi$ is smaller than the rest of the scales and they may depend on the geometry of the partitioning.
A closer look at the data indeed reveals additional contributions to mutual information which spoil the area law \eqref{eq. MI scaling}.
To see this we plot in Fig.~\ref{fig. chi2} the $\sqrt{\chi^2/\textrm{DOF}}$ of the fits, which were performed assuming relation Eq.~\eqref{eq. MI scaling}. Three regions of inverse temperatures can be easily defined. At small $\beta$ the Ansatz Eq.~\eqref{eq. MI scaling} describes all system sizes since $\sqrt{\chi^2/\textrm{DOF}} \approx 1$, suggesting that finite volume effects are smaller than the statistical uncertainties. A similar situation occurs for large $\beta$, where again the fit involved all available system sizes. Contrary, in the region close to the phase transition, where the correlation length $\xi$ is the largest, the fits have clearly bad quality,  $\sqrt{\chi^2/\textrm{DOF}} \gg 1$. This picture is further confirmed by the fact that the quality of the fit improves when we discard smaller system sizes, $L<L_{min}$, as shown for the strip partitioning in the inset.

The values of $\sqrt{\chi^2/\textrm{DOF}}$ close to $\beta_c\approx 0.44$ are particularly large for the chessboard partitioning. However, since the errors of the individual points are much smaller than for the rest of the partitionings (due to the simplified calculation of $Z(\v a)$ - see 
Appendix \ref{ap. chessboard}), we refrain from drawing conclusions about the size of finite-size effects in this partitioning compared to block partitionings.

\begin{figure*}
\begin{center}
\includegraphics[width=0.495\textwidth]{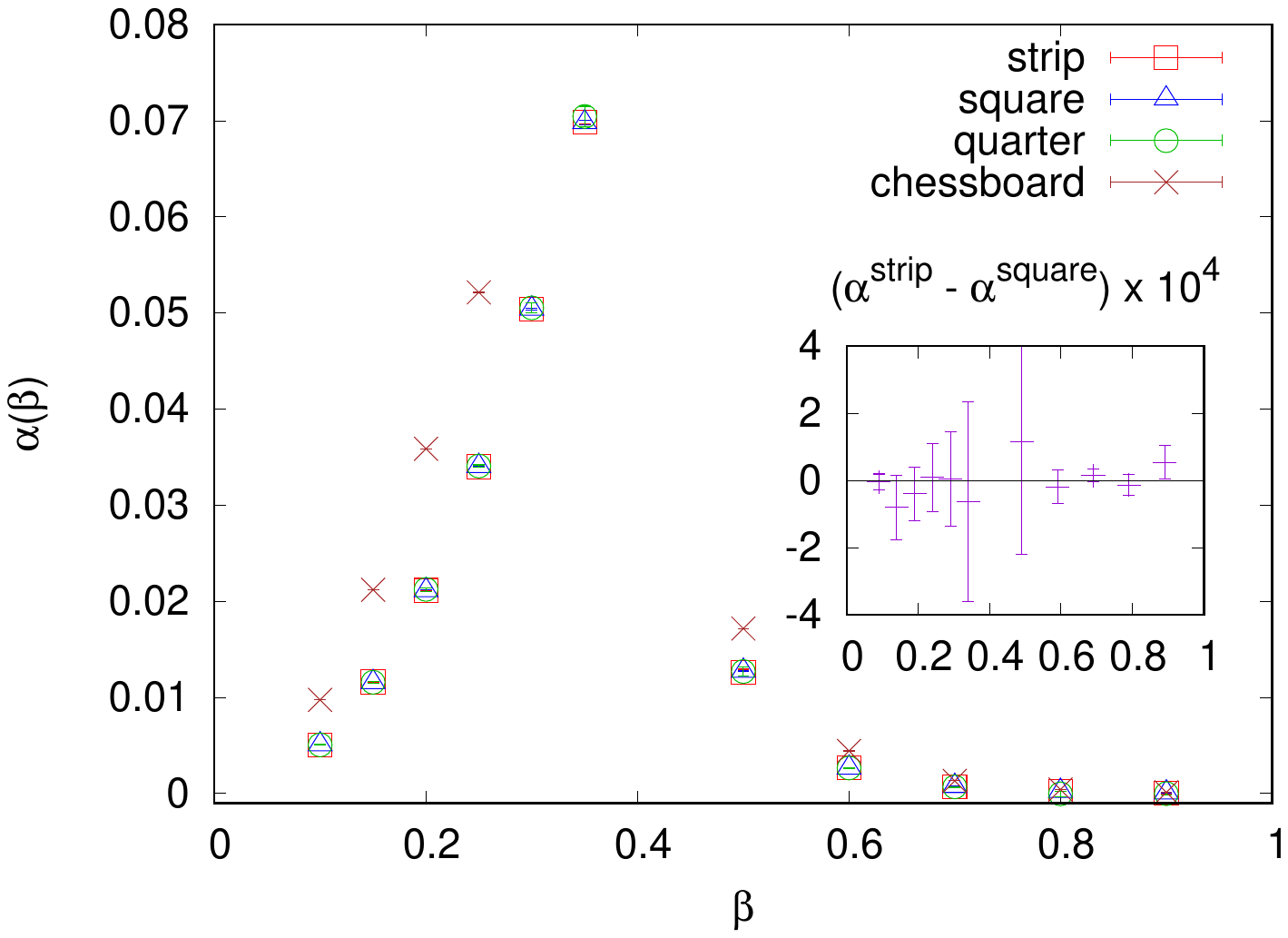}
\includegraphics[width=0.495\textwidth]{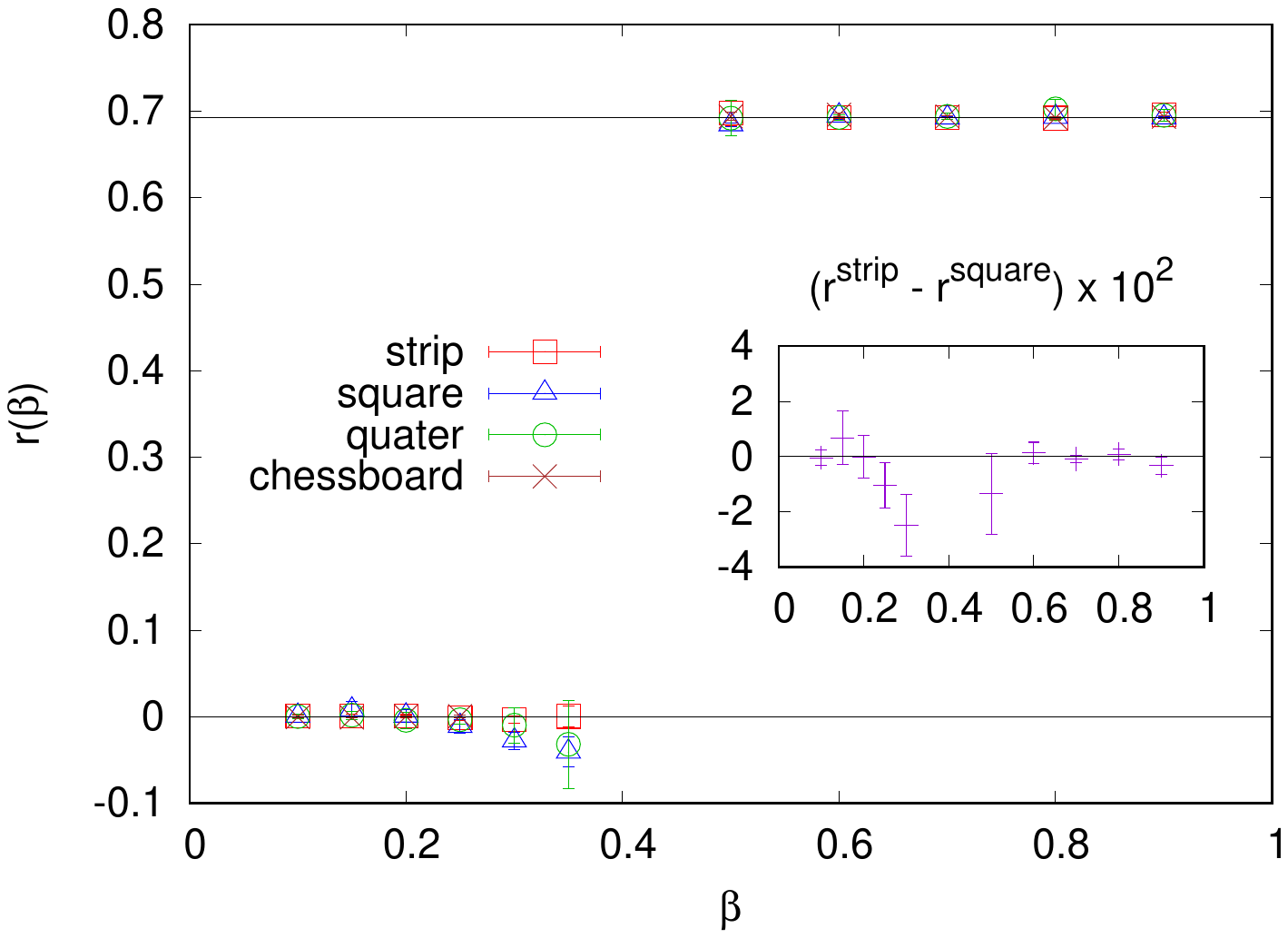}
\vspace{-1.0cm}
\caption{Coefficients $\alpha$ and $r$ for four geometries shown in Figure \ref{partition_geom} as a function of the inverse temperature $\beta$. The inset shows the difference between $\alpha$ and $r$ from strip and square geometries. Both geometries yield results compatible within their statistical and systematical uncertainties. \label{fig. main}}
\end{center}
\vspace{-0.90cm}
\end{figure*}

\subsection{Discussion of $\alpha(\beta)$ and $r(\beta)$ coefficients} 

For $\beta$ far from critical inverse temperature we can reliably describe our data with the area law Ansatz Eq.~\eqref{eq. MI scaling}. This allows us to extract the coefficients $\alpha$ and $r$ for the four different partitionings and compare them. We show the results in Fig.~\ref{fig. main}. In the main figures, we show the dependence on $\beta$ whereas in the insets we show the difference between the "strip" and "square" partitionings (the differences between other partitionings look similar). We have shown only the values which were obtained from fits with  $\sqrt{\chi^2/\textrm{DOF}} \lesssim 1$, as to be sure that the postulated dependence Eq.~\eqref{eq. MI scaling} is indeed reflected in the data. This means $\beta\le 0.35$ and $\beta\ge 0.5$ for block partitionings,  $\beta\le 0.25$ and $\beta\ge 0.5$ for chessboard. The uncertainties of data points in Fig.~\ref{fig. main} contain: i) the propagation of statistical uncertainties of $\hat{I}_{N,M}(\beta,L)$ through the extrapolation fits ($M \rightarrow \infty$ and the dependence on $L$) and ii) the systematic uncertainty of the fits estimated by taking the maximal difference of the outcomes when using different fit Ansatze or fit intervals. Both types of uncertainties were added in quadrature to obtain the final error bars shown in Fig.~\ref{fig. main}.

In the left panel, we show the coefficient $\alpha$ as a function of $\beta$. Qualitative behavior of $\alpha(\beta)$ seems to be universal: it goes to 0 at $\beta=0$ and $\beta\rightarrow \infty$ and rises around the critical temperature. The data clearly show that the chessboard partitioning yields different values than the three other possibilities which seem to be compatible with each other. The differences shown in the inset are indeed compatible with zero within their uncertainties. Therefore, we conclude that in this range of $\beta$ the partitioning does not influence the mutual information (when block partitionings are considered). In the right panel, we show the value of $r$. In that case, all four partitionings give the same result: 0 for $\beta<\beta_c$ and $\ln 2$ for $\beta>\beta_c$. \footnote{For $\beta \rightarrow \infty$ the space of states reduces to the two states -- one with all spins up and second with all spins down -- both having the same probability $1/2$. Plugging such probability distribution to Eq.~\eqref{eq. MI} one gets $I(\beta=\infty)=\ln 2$}

Comparing our results to the ones from Ref.~\cite{Lau_2013}, where the bond propagation and transfer matrix methods were used to calculate mutual information for cylinder-like geometry in the limit of infinite length of the cylinder, we may make several observations. First, we note that our $r(\beta)$ follows the behavior $r(\beta<\beta_c)=0$, and  $r(\beta>\beta_c)=\ln 2$ obtained in Ref.~\cite{Lau_2013} in the limit of infinite cylinder's circumference. The deviations from this behavior which we observe close to $\beta_c$ should be attributed to the finite size effects. Finite volume effects are also responsible for the fact that we cannot reproduce the value of $r(\beta_c)$ at the critical point, Ref.~\cite{Lau_2013}: $r(\beta=\beta_c)=0.2543925(5)$. When comes to the $\alpha$ coefficient, in Ref.~\cite{Lau_2013} only the critical value was calculated $\alpha(\beta=\beta_c)=0.37692626(7)$. The low-$\beta$ leading behavior of $\alpha$ was obtained in Ref.~\cite{2020PNAS..11730234N} using the partitioning which we called "strip": $\alpha(\beta)\approx \frac{1}{2} \beta^2 $ for $\beta \rightarrow 0$ and our results follow exactly this behavior.

\section{Summary} 

In this work, we have provided a numerical demonstration that the Shannon bipartite mutual information (MI) can be readily obtained from the Neural Importance Sampling (NIS) algorithm for the classical Ising model on a square lattice. Our approach allows for studies of different partitionings and we provided comparisons of the mutual information estimated for four geometries. We successfully exploited the hierarchical algorithm (HAN) to reach larger system sizes than achievable using standard Variational Autoregressive Networks (VAN).
The crucial property of our approach is that it provides unbiased results  (unlike the MICE~\cite{2020PNAS..11730234N} approach which is variational) assuming all modes of the target probability distribution can be probed.

It is important to note that the condition that the probability distribution modeled by the neural network contains all the modes of the target distribution is of central importance for the NIS approach (and in more general, for all MCMC methods). Obviously, when mode collapse is present and some part of the target distribution support is not sampled, the re-weighting procedure cannot compensate for the missing contributions to the partition function, which may lead to systematic bias. It is known that generative neural networks are sensitive to the mode collapse \cite{2019PhRvL.122h0602W,Hackett:2021idh,2023MLS&T...4a0501C,2022arXiv220801893I}. In the case of the Ising model considered here, where analytical results for the partition function are known, the possibility of mode collapse can be very much excluded by checking that the NIS and exact result agree.

We discussed the validity and universality of the area law for different partitionings. We found that at low and high temperatures the area law is satisfied whereas such an Ansatz does not describe our data in the vicinity of the phase transition, supposedly due to inherent finite volume effects. 

\section{Outlook}

Our proposal can be applied to many other spin systems with short- and long-ranged interactions, such as various classes of spin-glass. However, the main difficulty one encounters when dealing with spin-glass systems is the efficiency of training these complicated Boltzmann distributions, and in particular, the avoidance of mode collapse issues. As such, the applicability of our method depends on the progress in the neural network sampler's efficiency. Also, the long-ranged interactions prohibit the application of the HAN algorithm limiting, at the moment, the available system sizes down to $L\sim 30$.

We believe that by exploiting the Feynman path integral quantization prescription one may use the approach based on autoregressive networks to estimate also the entanglement entropy in quantum spin systems. In such a picture, a $D$ dimensional quantum system is described by a $D+1$ statistical system, where the machine learning enhanced Monte Carlo is applicable. In particular, thanks to the replica method \cite{Callan:1994py}, one can directly express the Renyi entropies in terms of partition functions \cite{Humeniuk:2012xg} readily obtainable in the NIS. In this approach systems in any space-time dimension can be studied, with the obvious limitation that performance is limited by the total number of sites in the system, hence reducing the available volumes in higher dimensions. Still, due to its straightforwardness, it should be seen as a valuable alternative to studying information-theoretic properties of one-, two-, or three-dimensional quantum systems.
In particular, the method based on path integral quantization can be considered as complementary to variational approaches based on autoregressive networks such as Refs.~\cite{2020PhRvR...2b3358H,2020PhRvA.102f2413W} where the approximation of the ground state wave function is constructed. First, with an ergodic sampling it provides unbiased results (see discussion in Appendix \ref{app_MICE_comp}); and second, it gives access to the entanglement entropy for thermal states \cite{2004JSMTE..06..002C,Itou:2015cyu}.

One should also observe that our proposal in principle can work unaltered for systems with continuous degrees of freedom. Neural generative networks have been already discussed in the context of $\phi^4$ classical field theory \cite{PhysRevLett.126.032001} as well as $U(1)$, $SU(N)$ \cite{Abbott:2022zhs}, and the Schwinger model gauge theories \cite{Albergo:2022qfi}. In all these cases, one would introduce a conditional normalizing flow or another proposal that would give access to conditional probabilities (see for example Ref.~\cite{Wang_2022}). For instance, a hierarchical construction similar to the HAN algorithm employing normalizing flows could be used to simulate the $\phi^4$ classical field theory. In this way, conditional probabilities would be naturally introduced and could be used to calculate the mutual information for some specific partitioning. The combination of the proposed method of measuring information-theoretic quantities with the recent advancements \cite{Abbott:2022hkm} in the machine learning enhanced algorithms for simulating four-dimensional Lattice Quantum Chromodynamics (LQCD) can open new ways of investigating this phenomenologically important theory, for instance by studying quantum correlations and entanglement of the QCD vacuum.

\section*{Acknowledgments}
Computer time allocation 'plgng' on the Prometheus and ARES supercomputers hosted by AGH Cyfronet in Krak\'{o}w, Poland was used through the Polish PLGRID consortium. T.S. kindly acknowledges the support of the Polish National Science Center (NCN) Grants No.\,2019/32/C/ST2/00202 and 2021/43/D/ST2/03375 and support of the Faculty of Physics, Astronomy and Applied Computer Science, Jagiellonian University Grant No.~LM/23/ST. P.K. acknowledge that this research was partially funded by the Priority Research Area Digiworld under the program Excellence Initiative – Research University at the Jagiellonian University in Kraków. P.K. and T.S thank Alberto Ramos and the University of Valencia for their hospitality during the stay when part of this work was performed and discussed. We also acknowledge very fruitful discussions with Leszek Hadasz on entanglement entropy.




\appendix

\section{Details of the method}
\label{ap. systematics}

In order to calculate the mutual information according to the definition Eq.~\eqref{eq. MI4} we need to estimate the averages $\langle \ldots \rangle_{q_{\theta}}$ which we do using the Monte Carlo approach,
\begin{equation}
    \langle \mathcal{O} \rangle_{q_{\theta}} \approx \frac{1}{N}\sum_{i=1}^N \mathcal{O}(\mathbf{s}_i)\qquad \mathbf{s}_i \sim q_{\theta},
\end{equation}
where the configurations $\v s_i$ are drawn from the distribution $q_{\theta}$. In particular, the partition function can be approximated as
\begin{multline}
    Z=\sum_{\v a, \v b}\qt(\v a, \v b) \frac{e^{-\beta E(\v a, \v b)}}{\qt(\v a, \v b)}\equiv\avg{\hat w(\mathbf{a, b})}_{q_{\theta}(\v a, \v b)} \\
    \approx  \frac{1}{N}\sum_{i=1}^N \hat{w}(\v a_i, \v b_i)\equiv \hat Z_N,   \qquad (\v a_i, \v b_i) \sim q_{\theta}.
    \label{def_Z_N}
\end{multline}
In a similar way,
\begin{equation}
    Z(\v a) =
     \sum_{\v b}\qt(\v b|\v a) \frac{e^{-\beta E(\v a, \v b)}}{\qt(\v b|\v a)}\equiv\langle  \hat w(\mathbf{b|a}) \rangle_{q_{\theta}(\mathbf{b|a})},
\end{equation}
where we introduced $\hat w(\mathbf{b|a})= e^{-\beta E(\mathbf{a}, \mathbf{b}_i)}/q_{\theta}(\mathbf{b}|\mathbf{a})$ and $\langle \ldots \rangle_{q_{\theta}(\mathbf{b|a})}$ is an average over the conditional probability. This can be approximated as
\begin{equation}
    Z(\v a) \approx \hat Z_M(\mathbf{a}) = \frac{1}{M}\sum_{i=1}^M \frac{e^{-\beta E(\mathbf{a}, \mathbf{b}_i)}}{q_{\theta}(\mathbf{b}_i|\mathbf{a})},\qquad\mathbf{b}_i \sim q_{\theta}(\mathbf{b}|\mathbf{a}),
    \label{def_Z_M_a}
\end{equation}
where we have denoted by $M$ the number of configurations used to estimate that average. In general, $M$ is independent of $N$, for practical reasons we always take $M < N$. To calculate $Z(\v b)$ we  need $\qt(\v a|\v b)$ which is not readily available. Therefore, we define a new distribution $\qtt$ obtained by feeding a permutation of spins that  swaps $\v a$ and $\v b$ into $\qt$. For VAN this  is obtained by reversing the order of spins
\begin{equation}
\qtt(s^1,s^2,\ldots,s^{n_A+n_B})\equiv \qt(s^{n_A+n_B},s^{n_A+n_B-1},\ldots,s^{1}).
\end{equation}
We can use $\qtt$ together with factorisation Eq.~\eqref{eq:factorisation} to obtain $\qtt(\v a|\v b)$ and $Z(\v b)$,
\begin{equation}
    Z(\v b)\approx\hat Z_M(\mathbf{b}) = \frac{1}{M}\sum_{i=1}^M \frac{e^{-\beta E(\mathbf{a}_i, \mathbf{b})}}{\qtt(\mathbf{a}_i|\mathbf{b})},\qquad\mathbf{a}_i \sim \qtt(\mathbf{a}|\mathbf{b}).
    \label{def_Z_M_b}
\end{equation}
Please note that in general $\qtt(\v s)\neq \qt(\v s)$ but for a well-trained network these two distributions should be close. Furthermore, the approximation Eq.~\eqref{def_Z_M_b} is valid for any $\qtt$. 

For the remaining terms we use the standard way to calculate the average, i.e.~the mean energy is
\begin{multline}
\frac{1}{Z}\langle  \hat w(\v a, \v b) E(\v a, \v b) \rangle_{q_{\theta}(\v a, \v b)} \\ \approx \frac{1}{N\hat Z_N}\sum_{i=1}^N \hat{w}(\v a_i, \v b_i) E(\v a_i, \v b_i)  \qquad (\v a_i, \v b_i) \sim q_{\theta}.
\label{mean_energy_calculation}
\end{multline}
In that case, the statistical uncertainty of the result is governed by the square root of the number of samples generated $N$. The mean logarithm of $Z(\v a)$ estimated as
\begin{multline}
\frac{1}{Z} \langle \hat w(\v a, \v b) \log Z(\v a) \rangle_{q_{\theta}(\v a, \v b)}\\ 
\approx \frac{1}{N\hat Z_N}\sum_{i=1}^N \hat{w}(\v a_i, \v b_i) \log \hat Z_M(\v a_i),  \qquad (\v a_i, \v b_i) \sim q_{\theta}
\label{logZa_estim}
\end{multline}
and analogously for $\log Z(\v b)$. 

The procedure of calculating Eq.~\eqref{logZa_estim} is the following: i) we draw $N$ configurations of the entire system using the probability distribution $q_{\theta}$ encoded by the neural network, ii) we estimate the full partition function $\hat Z_N$ according to (\ref{def_Z_N}), iii) for each of the $N$ configurations we froze the subsystem $\v a$ and generate additional $M-1$ configurations with random part $\v b$ using the conditional probability $q_{\theta}(\mathbf{b}|\mathbf{a})$, iv) for each of the $N$ configurations we calculate $\hat Z_M(\mathbf{a})$ using Eq.~\eqref{def_Z_M_a} using the $M$ configurations with the same frozen part $\v a$, v) we calculate the average over $N$ configurations of the logarithm of $\hat Z_M(\mathbf{a})$ according to Eq.~\eqref{logZa_estim}. Steps iii)-v) can be analogously applied to calculate $\langle w \log Z(\v b) \rangle_{q_{\theta}}$.

Adding up the terms according to Eq.~\eqref{eq. MI4} we obtain the estimator of the mutual information $\hat I_{N,M}$ which yields the exact value in the following infinite-statistics limit:
\begin{equation}
    I=\lim_{N\rightarrow \infty}\lim_{M\rightarrow \infty} \hat I_{N,M}.
\end{equation}
At finite $N$ and $M$ our estimator $\hat I_{N,M}$ is biased due to the nonlinearity of the $\log$ function. As was shown in the Appendix of Ref.~\cite{Nicoli:2020njz}, $\log\hat Z_N$ has a  bias due to the finite value of $N$ which is given by
\begin{multline}
    \mathcal{B}[\log\hat Z_N]=  -\frac{1}{2N} \frac{\langle  \hat w(\mathbf{a, b})^2 \rangle_{q_{\theta}(\v a, \v b)} -\langle  \hat w(\mathbf{a, b}) \rangle_{q_{\theta}(\v a, \v b)}^2 }{\langle  \hat w(\mathbf{a, b}) \rangle_{q_{\theta}(\v a, \v b)}^2} \\+\mathcal{O} \left(  \frac{1}{N^2}\right), 
    \label{bias_Z_N}
\end{multline}
where the numerator is equal to the variance of $Z$. This is a systematical bias of the observable and would be zero for a perfectly trained network, i.e. when $q_{\theta}=p \Leftrightarrow \hat w(\mathbf{a, b})=const$.  For $N$ sufficiently large, this bias can be neglected as it decreases with $1/N$ and is usually smaller than the statistical noise which decreases only as $\sim 1/\sqrt{N}$. In practical terms, with our statistics of $N\approx 10^6$ we are always working in this regime. Also, $\log \hat Z_M(\v a)$ is affected by a similar bias for a finite value of $M$. Repeating the calculation for that observable, one can obtain in analogy to Eq.~\eqref{bias_Z_N},
\begin{multline}
    \mathcal{B}[\log \hat Z_M(\v a)]=
    -\frac{1}{2M} \frac{\langle  \hat w(\mathbf{b|a})^2 \rangle_{q_{\theta}(\mathbf{b|a})} -\langle  \hat w(\mathbf{b|a}) \rangle_{q_{\theta}(\mathbf{b|a})}^2 }{\langle  \hat w(\mathbf{b|a}) \rangle_{q_{\theta}(\mathbf{b|a})}^2}\\ +\mathcal{O} \left(  \frac{1}{M^2}\right),
\end{multline}

\begin{figure}
\begin{center}
\includegraphics[width=0.495\textwidth]{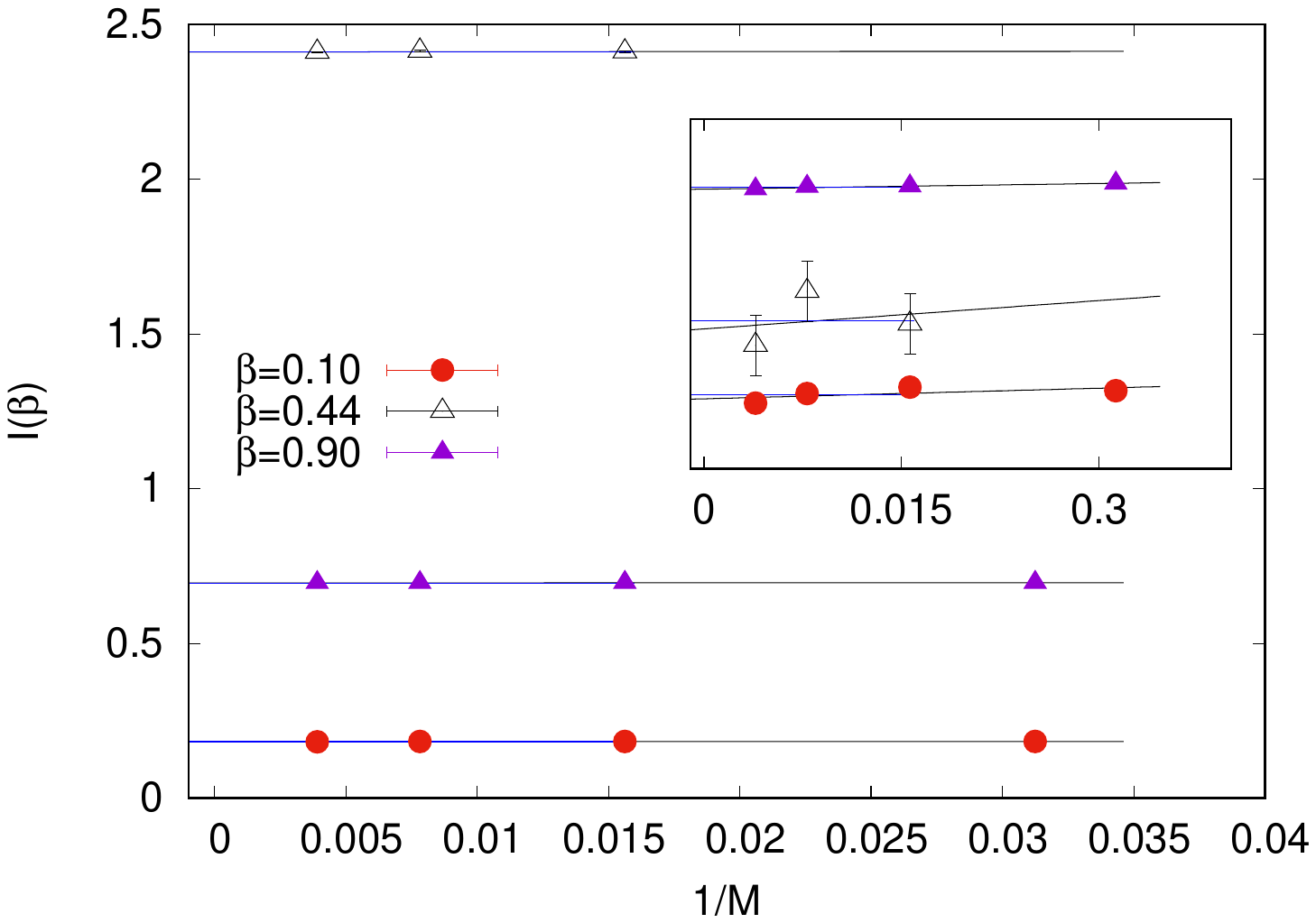}
\caption{Dependence on $M$ at $L=18$ for strip partitioning for three representative values of $\beta=0.1, 0.44$ and $0.9$. Two extrapolations are shown: constant and linear. Their difference is taken as a systematic uncertainty. The inset shows the same data shifted by a constant amount to show the details.  \label{fig. dependence on M}}
\end{center}
\end{figure}

In the practical implementation, it is difficult to afford $M \sim N \sim 10^6$ so typically $M \sim 10^2 \ll N$. Therefore, 
neglecting $\mathcal{B}[\log\hat Z_N]$, the final systematic bias of $\hat I_{N,M}$ is given by
\begin{multline}
     \mathcal{B}[\hat I_{N,M}]=\frac{1}{2Z M} \bigg\langle \hat w(\mathbf{a, b}) \times \\  \left( \frac{\langle  \hat w(\mathbf{b|a})^2 \rangle_{q_{\theta}(\mathbf{b|a})} -\langle  \hat w(\mathbf{b|a}) \rangle_{q_{\theta}(\mathbf{b|a})}^2 }{\langle  \hat w(\mathbf{b|a}) \rangle_{q_{\theta}(\mathbf{b|a})}^2} + 
     \v a \leftrightarrow \v b \right) \bigg\rangle_{q_{\theta}}
     \\ +\mathcal{O} \left(  \frac{1}{M^2}\right).
     \label{bias_INM}
\end{multline}
We expect that a positive bias may affect $I$ which decreases as $\sim 1/M$ for large enough $M$. In our calculation, we take $M=64,128,256$ and in some cases, $M=512$ and $M=1024$, and perform an extrapolation to $M=\infty$. We attach a systematic uncertainty in that step by comparing the result of a constant extrapolation to the data at the two largest values of $M$ with the linear $a + b^2/M$ extrapolation to the data at the three largest values of $M$. In fact, in most cases, we observe that $\hat I_{N, M}$ is equal within statistical errors for all values of $M$. We provide representative examples of such extrapolations in Fig.~\ref{fig. dependence on M}.

\section{Neural network architectures} 
\label{ap. neural nets}

In this work, we use two algorithms employing autoregressive neural networks: Variational Autoregressive Networks (VAN) \cite{2019PhRvL.122h0602W} and their modification called Hierarchical Autoregressive Networks (HAN) \cite{Bialas:2022qbs}. Both architectures provide access to the conditional probabilities Eq.~\eqref{eq:factorisation}.
The training of the neural networks consists in generating $N_{batch}$ spin configurations $\{\v s_1 \ldots \v s_{N_{batch}} \}$ from the probability distribution $q_{\theta}$ currently encoded in the neural weights and calculating the variational estimate of the free energy:
\begin{equation}
F_q= \frac{1}{\beta}\sum_{k=1}^{N_{batch}} q_\theta(\mathbf{s}_k) \left[\beta H(\mathbf{s}_k)+\log q_\theta(\mathbf{s}_k) \right],
\label{F_q_def}
\end{equation}
which, up to an additive constant, corresponds to the backward Kullback-Leibler divergence between $q_{\theta}$ and $p$. With this loss function, the weights $\theta$ are updated according to the gradient backpropagation algorithm with ADAM optimizer \cite{ADAM}.

We use dense (fully connected) neural networks with two layers. The autoregressive property is enforced by multiplying half of the weights by 0. The neural network for the VAN approach is constructed as:
\begin{equation}
    \mathcal{N}_\theta=\sigma \circ l_2  \circ \textrm{ReLu} \circ l_1,
\end{equation}
where layer $l$ acts on the input vector $x\in \mathbb{R}^n$ in the following way
\begin{equation}
    y_i = \sum_{j<i} W_{ij} x_j +b_i, \ \textrm{for} \ y=l(x). 
    \label{layer_function}
\end{equation}
The autoregressive property of the neural network is guaranteed by the fact that the above sum runs only up to $i-1$.

The activation functions act pointwise on their arguments. We use two types of functions: the rectifier $\textrm{ReLu}(x)=\textrm{max}(0,x)$ and the sigmoid $\sigma(x) = 1/(1+e^{-x})$. Weights $W_{ij}$'s and biases $b_i$'s are parameters of the neural network and collectively denoted as $\theta$. 

The neural network acts on a spin configuration $\{s^1,\ldots s^{L^2} \}$ as follows: $\mathcal{N}_\theta(\{s^1,\ldots s^{L^2} \})=\{\hat s^1,\ldots \hat s^{L^2} \}$, where each $\hat s^i$ is interpreted as the conditional probability of spin $i$-th to be pointed up given all the previous $i-1$ spins were fixed: $ q_\theta(s^i=+1|s^{i-1},\ldots,s^1)=\hat s^i$. The autoregressive condition Eq.~(\ref{layer_function}) guarantees that the conditional probability of $i$-th spin depends only on the previous $i-1$ spins.

In the VAN approach, the conditional probabilities of all $L^2$ spins are generated by a single neural network $\mathcal{N}_\theta$ described above. It has $L^2$ input neurons and $L^2$ output neurons. To fix all the $L^2$ spins, one needs to evoke the neural network $L^2$ times: at $i$-th invocation one calculates $\hat s_i$ which is then used to draw a value of $s_i$. This leads to an unfavorable $\sim L^6$ rise in the numerical cost of the generation of samples (which is the main numerical cost of the algorithm). Additionally, the larger the system is, the more configurations are needed to train the network to given quality (measured for example by the Effective Sample Size (ESS) \cite{Liu}). This makes the effective numerical cost grow even faster than $\sim L^6$ \cite{DelDebbio:2021qwf}.

\begin{figure}
    \centering
    \includegraphics[width=3.5cm]{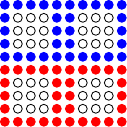}\kern5mm\includegraphics[width=3.5cm]{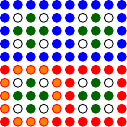}
    \caption{Sketch of the partitioning in HAN of a $L \times L=10 \times 10$ lattice. Border spins are shown in blue and red and the first set of interior spins in green. The probabilities of the green spins depend conditionally on all spins surrounding them (colored orange for one set of interior spins). \label{fig:sketch}}
    \label{fig:han}
\end{figure}

To mitigate the above-mentioned problems we use a version of the HAN algorithm proposed in Ref.~\cite{Bialas:2022qbs} which introduces a specific enumeration of the spins (see Figure~\ref{fig:han}): we first fix the frames (denoted in blue and red) which surround all the remaining spins, then iteratively we fix the "crosses" inside the frames (green crosses on Figure) to end up with single spins. There are several advantages of this division which come from the fact that, for nearest-neighbor interactions, the probability of a group of spins depends conditionally on the values of spins on a closed contour enclosing that set (result known also as the Hammersley-Clifford theorem in the literature \cite{Hammersley-Clifford, Clifford90markovrandom}). First, instead of one network for the whole system, one can use several smaller networks which fix the spins at a given level of hierarchy (levels of hierarchy in our Figure~\ref{fig:han} are denoted by different colors\footnote{Note that, in principle, red and blue spins could be treated as belonging to the same level of the HAN hierarchy. However, here we explicitly distinguish them because we need to conditionally sample red spins based on blue spins and calculate the conditional probabilities needed for the mutual information observable. Therefore, red spins have a separate network than the blue ones.}). The networks generating the spins in the crosses depend conditionally only on the surrounding spins (denoted by orange for one cross). Therefore those networks are much smaller than the single network in the VAN approach, hence the numerical cost is significantly reduced. What is more, at a given level of the hierarchy, the crosses can be generated in parallel. As was shown in \cite{Bialas:2022qbs}, the numerical cost of HAN is much smaller than VAN, it scales\footnote{We note that the $\sim L^6$ scaling for the VAN approach is due to the sampling procedure: the neural network has to be evaluated $L^2$ times to fix all the spins and naively each evaluation requires $\sim L^4$ floating point operation as the network has $L^2$ neurons per layer. However, this counting assumes that operations using all neural network weights are done during each evaluation of the network. 
In a more efficient implementation
of autoregressive networks, one would perform multiplication with only the weights which are necessary to fix the given spin. It is easy to check that the scaling of the numerical cost of the VAN algorithm is then $\sim L^4$ \cite{2016arXiv160502226U}. By the same argument, the scaling for the HAN algorithm with optimal implementation is $L^2$. This is because the first neural network in the HAN hierarchy describes the spins at the border of the lattice, which has size $\sim L$. } 
as $\sim L^3$.
Smaller networks are also easier to train so with the same number of epochs one reaches a much higher ESS with HAN.

In practice, we are limited to $L<30$ in VAN simulations, whereas with HAN we can reach sizes $L=66$ (or even $L=130$ for some temperatures). On the other hand, as the crosses in HAN can only have specific numbers of spins (in order to close the recurrence), this algorithm can be used for simulations with $L=10,18,34,66,130,\ldots$ whereas VAN can be applied to any $L$ value. Also, the VAN is much more elastic concerning the possible divisions into $\v a$ and $\v b$ subsystems: with the proper enumeration of the spins {\it any} division is possible in VAN. In HAN, with the implementation described here, only strip and square partitionings are possible. Each division that requires a specific enumeration of the spins requires also new training. 

In this manuscript, we used both VAN and HAN algorithms: the latter was used to obtain mutual information for  $L=10,18,34,66,130$, with strip and square geometries. All other values of $I$ were obtained using VAN. In order to show the numerical cost of the method we provide the runtime of the $L=66$ system simulated with the HAN algorithm (one of the largest simulated): we trained the hierarchy of neural networks for 220000 epochs, which took 110 h on NVIDIA V100 graphic card to reach the ESS of $0.508$. The model had 2 200 000 parameters. The consecutive measurement of the mutual information for $M=256$ took 20 h. After repeating such measurements for $M=128$ (10h of running) and $M=64$ (5h of running) we performed the extrapolation in $M$. This allowed us to achieve a total uncertainty of $I$ of $0.15 \%$ at $\beta=0.44$ which is the worst case.

\section{Chessboard factorization} 
\label{ap. chessboard}

The consequence of the Hammersley-Clifford theorem is that once all four nearest-neighbors $n(s^i)$ of the spin $s^i$ are fixed, its probability is given by:
\begin{align}
p_{chess}(s^i=\pm 1, n(s^i))=  \left[ 1+\exp \left(\mp\, 2\beta \mathcal S_i  \right) \right]^{-1},
\label{chess_prob}
\end{align}
where $\mathcal S_i=\sum_{\ j \in n(s^i) } s^j$ and $j$ runs over four nearest-neighbors of spin $i$.

In \cite{Bialas:2021bei} we proposed to use this property to reduce the number of spins that need to be generated by the network by a factor of 2. The idea is to divide the spin configuration into subsystems according to the chessboard pattern: spins at white fields are fixed by the network using the usual VAN algorithm (we call them $\v a$). The other half of the system (called $\v b$) can be drawn from probabilities Eq.~\eqref{chess_prob}.

Such factorization makes calculating the mutual information using chessboard partitioning particularly simple. We note that due to symmetry reasons: $\langle \hat w \log Z(\v a) \rangle_{q_{\theta}} = \langle \hat w \log Z(\v b) \rangle_{q_{\theta}}$, hence only the former needs to be calculated. For this purpose, we write
\begin{align}
E(\v a, \v b)= -\sum_{\langle i,j \rangle} s^i s^j = -\sum_{i \in \v b}s^i \mathcal S_i(\v a), 
\end{align}
where we explicitly denoted that  $\mathcal S_i$ for $i \in \v b$ depends only on the subsystem $\v a$ (since any spin from $\v b$ has nearest-neighbors only from $\v a$). Then
\begin{multline}
\log Z(\v a)= \log \sum_{\v b} e^{\beta \sum_{i \in \v b}s^i \mathcal S_i(\v a) } = \\
=\log \prod_{i \in \v b}\left( \sum_{ s^i=\{-1,1\}  } e^{\beta s^i \mathcal S_i(\v a) } \right)=  \sum_{i \in \v b} \log \left[ 2\cosh \beta \mathcal S_i(\v a) \right],
\end{multline}
which means that $\log Z(\v a)$ can be exactly calculated for any $\v a$. In other words, in the chessboard division, there is no need for Eq.~\eqref{def_Z_M_a}, the observable $\log Z(\v a)$ can be exactly determined for any configuration and  is not biased. This means that we need much less statistics to determine mutual information.

\section{Symmetries}
\label{ap. symmetries}

In order to obtain good quality training of the autoregressive neural network, it is crucial to impose global symmetries of the system through the symmetrization of the loss function \cite{2019PhRvL.122h0602W,Bialas:2022qbs, Bialas:2021bei,Bialas:2022bdl}. This can be achieved by defining a symmetrized probability for each generated configuration
\begin{equation}
    \bar{q}_{\theta}(\mathbf{s}) = \frac{1}{S}\sum_{i=1}^S q_{\theta}( h_i(\mathbf{s})),
\end{equation}
where $h_i$, $i=1,\dots,S$ are the symmetry operators defined as transformations of the configuration space which keep the energy unchanged. This symmetrized probability replaces $q_\theta$ in the Kullback--Leibler (KL) loss function:
\begin{align}
    \bar{D}_\textrm{KL} (q_\theta | p) &= \sum_{k=1}^{N_{\textrm{batch}}} \bar{q}_\theta(\mathbf{s}_k) \, \log \left(\frac{\bar{q}_\theta(\mathbf{s}_k)}{p(\mathbf{s}_k)}\right).
    \label{eq:KL_loss_sym} 
\end{align}
and in the definition of importance ratios Eq.~\eqref{importance_ratio_def}. See Appendix B of Ref.~\cite{Bialas:2022bdl} for a more detailed discussion.

In case of $Z(\v a)$ or $Z(\v b)$, the calculation of the conditional probabilities $q_{\theta}(\mathbf{b}|\mathbf{a})$ and $q_{\theta}(\mathbf{a}|\mathbf{b})$ is needed. Since the partitioning itself breaks the symmetry, symmetrization is not possible. Hence, also at the level of conditional probabilities, one cannot impose any symmetry and so one is left with the non-symmetrized version of $q_\theta$.

\section{Numerical calculation of $I$ for $L=6$}

\begin{figure}
\begin{center}
\includegraphics[width=0.495\textwidth]{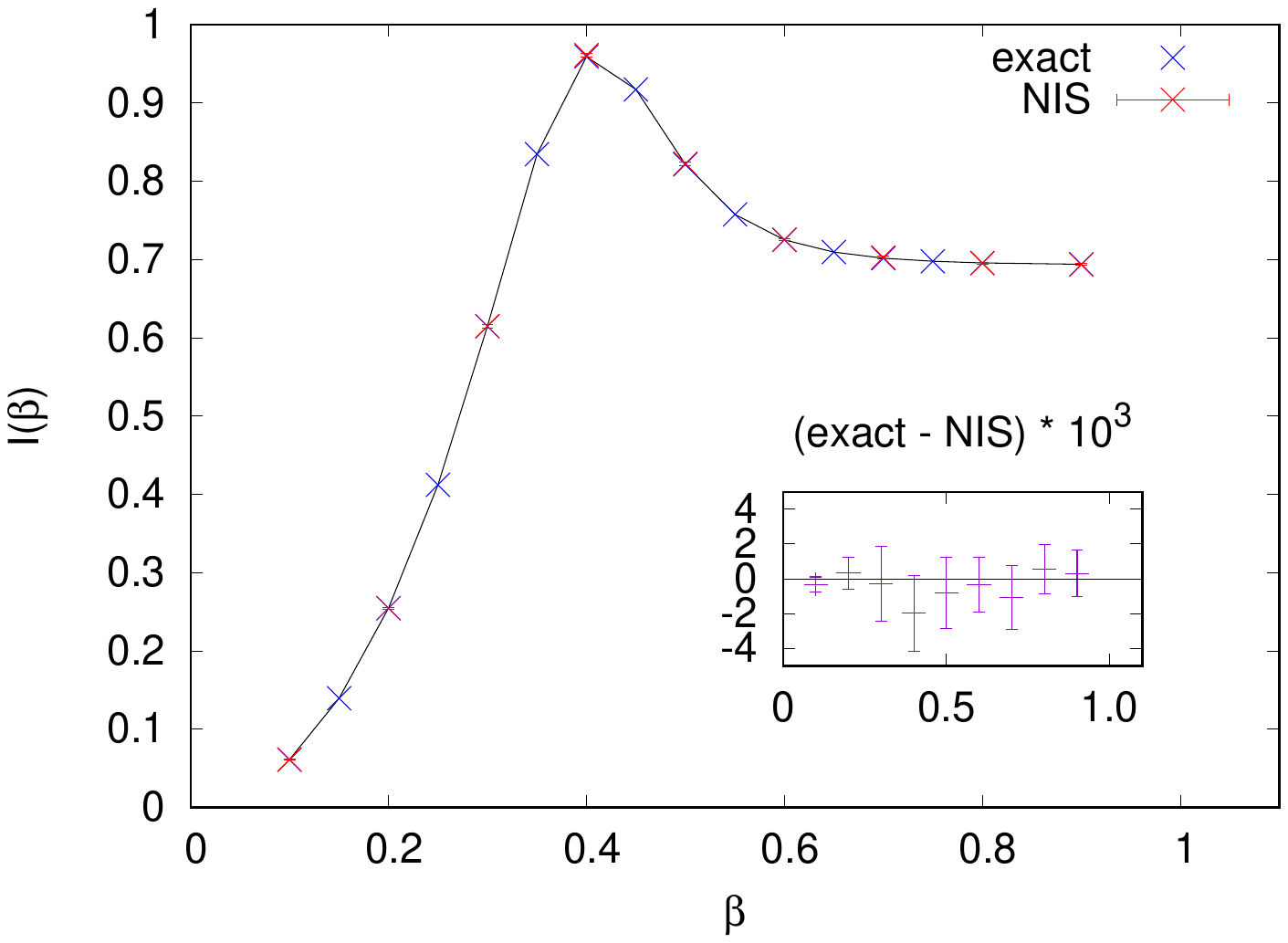}
\caption{Comparison of $I$ for strip partitioning calculated exactly from Eq.~\eqref{eq. MI2} and with the NIS approach through Eq.~\eqref{eq. MI4} for $L=6$. NIS data has statistical uncertainties which are smaller than the symbol size. Differences between the exact values and NIS are magnified in the inset.\label{fig. exact vs NIS}}
\end{center}
\end{figure}

For very small lattice sizes the number of states is small enough to calculate the mutual information directly from Eq.~\eqref{eq. MI2}. We were able to perform this calculation for $L=6$, where the number of states is $2^{36}$. Comparison with the results using VAN is shown in Fig.~ \ref{fig. exact vs NIS}. We found perfect agreement which we show by plotting the difference between the two methods in the inset.

\section{Comparison with MICE~\cite{2020PNAS..11730234N} }
\label{app_MICE_comp}

We now discuss the differences between NIS and MICE \cite{2020PNAS..11730234N} methods.

In the MICE approach, one uses the fact that mutual information can be treated as the KL divergence~\cite{pmlr-v80-belghazi18a} which in turn satisfies the Donsker and Varadhan theorem \cite{Donsker}: the $I$ is an upper bound of some variable $\cal M_\theta$ over the set of functions parameterized by a neural network. One then trains the network to maximize $\cal M_\theta$. With such construction, the MICE method is variational: it provides an approximation of $I$ which is in general smaller than the true value. However, as is typical in variational approaches, without knowing the exact result one cannot deduce the systematic uncertainty of the approximation. Applying  MICE to the 2D Ising model the Authors of Ref.~\cite{2020PNAS..11730234N} obtained the global entropy with an accuracy below 5\%. The bias for the MI, for which analytic values are not known, may be larger.

The NIS approach that we discuss in this manuscript circumvents the above-mentioned inaccuracy of the variational approach using a re-weighting procedure: due to the fact that the VAN/HAN procedures use explicit $q_\theta$ probabilities, one can remove the difference between $p$ and $q_\theta$ by calculating the weights $\hat w( a ,  b)$ and $\hat w( a | b)$ - as discussed in Appendix \ref{ap. systematics}, and correcting the final outcome. Therefore, our method provides, in the limit of large statistics, the exact result assuming the ergodicity condition of the algorithm is satisfied. The combined statistical and systematic uncertainty of MI obtained with NIS is less than 0.1\%.  

\bibliographystyle{ieeetr}
\bibliography{references2}

\end{document}